\documentclass[pra,twocolumn,amsmath,amssymb,showpacs,nofootinbib,longbibliography]{revtex4-1}

\newcommand{\Jnature}{Nature (London)}
\newcommand{\Jnatphys}{Nat. Phys.}

\newcommand{\JSciRep}{Sci. Rep.}
\newcommand{\Jscience}{Science}

\newcommand{\Jprl}{Phys. Rev. Lett.}

\newcommand{\Jpra}{Phys. Rev. A}
\newcommand{\Jprb}{Phys. Rev. B}

\newcommand{\Jrmp}{Rev. Mod. Phys.}

\newcommand{\Jnjp}{New J. Phys.}

\usepackage[english]{babel}
\usepackage{latexsym}
\usepackage{graphics}
\usepackage{subfigure}
\usepackage{epsfig}
\usepackage{color}
\usepackage{hyperref}
\usepackage{braket} 
\usepackage[T1]{fontenc}
\usepackage[latin9]{inputenc}
\usepackage{amstext}
\usepackage{amssymb}
\usepackage{graphicx}
\usepackage{esint}
\usepackage{braket}
\usepackage{babel}
\usepackage{amsmath}
\usepackage{bbm}
\pdfoutput=1
\hypersetup{
colorlinks=true,
citecolor=blue,
linkcolor=red,
urlcolor=black
}
\usepackage[normalem]{ulem}

\usepackage{times}

\newcommand\avg[1]{\langle #1 \rangle}
\renewcommand{\t}[1]{\text{#1}}
\renewcommand{\d}{\mathrm{d}}
\renewcommand{\vec}[1]{\boldsymbol{#1}}
\newcommand{\mc}[1]{\mathcal{#1}}
\newcommand{\e}{\mathrm{e}}
\renewcommand{\d}{\mathrm{d}}
\newcommand{\ie}{\textit{i.e.}~}

\definecolor{orange}{rgb}{0.85,0.5,0.2}

\usepackage{soul}


\begin{document}

\title{Local quench spectroscopy of many-body quantum systems}

\author{L. Villa}
\affiliation{CPHT, CNRS, Institut Polytechnique de Paris, Route de Saclay, 91128 Palaiseau, France}

\author{J. Despres}
\affiliation{CPHT, CNRS, Institut Polytechnique de Paris, Route de Saclay, 91128 Palaiseau, France}

\author{S. J. Thomson}
\affiliation{CPHT, CNRS, Institut Polytechnique de Paris, Route de Saclay, 91128 Palaiseau, France}

\author{L. Sanchez-Palencia}
\affiliation{CPHT, CNRS, Institut Polytechnique de Paris, Route de Saclay, 91128 Palaiseau, France}
\date{\today}

\begin{abstract}
Quench spectroscopy is a relatively new method which enables the investigation of spectral properties of many-body quantum systems by monitoring the out-of-equilibrium dynamics of real-space observables after a quench. So far the approach has been devised for global quenches or using local engineering of momentum-resolved excitations. Here, we extend the quench spectroscopy method to local quenches. We show that it allows us to extract quantitative information about global properties of the system, and in particular the elementary excitation spectrum. Using state-of-the-art numerical methods, we simulate the out-of-equilibrium dynamics of a variety of quantum systems following various local quench protocols and demonstrate a general scheme for designing an appropriate local quench protocol for any chosen model. We provide detailed examples of how the local quench protocol can be realised in realistic current generation experiments, including ultracold atomic gases and trapped ion systems.
\end{abstract}

\maketitle
\section{Introduction}
Quantum many-body systems driven out of equilibrium exhibit a diverse range of behaviours which have stimulated much activity in recent years. A generic quantum system subject to a sudden quench is expected to relax to a steady state in thermal equilibrium, provided integrability or disorder does not break ergodicity. Beyond the asymptotic state, the manner in which the system approaches thermal equilibrium gives us information about its microscopic properties.
For instance, the dynamics of correlation functions contain information on how quantum information propagates throughout the system, which has prompted both experimental~\cite{cheneau2012light,fukuhara2013quantum,fukuhara2013microscopic,richerme2014non,jurcevic2014} and theoretical~\cite{manmana2009time,barmettler2012propagation,hauke2013spread,eisert2013breakdown,krutitsky2014propagation,carleo2014light,bonnes2014light,andraschko2015propagation,
cevolani2016spreading,buyskikh2016entanglement,fitzpatrick2018light,cevolani2018universal,frerot2018multispeed,despres2019twofold} investigations.
The question of whether further information can be extracted from quench dynamics is now attracting growing attention.

A first step towards extracting such information from the non equilibrium dynamics was taken by Lieb and Robinson, who demonstrated the existence of a bound for the correlation spreading of local operators in short-range models~\cite{lieb1972finite}.
A physical interpretation of this result was later developed using a semi classical quasiparticle picture~\cite{calabrese2006time}, which enabled generalisations to other models, including long-range systems~\cite{hauke2013spread,cevolani2015,frerot2018multispeed,cevolani2018universal,vanderstraeten2018quasiparticles}. The time-evolution of correlation functions displays a causal-cone-like structure in which the behavior close to the edges yields information about the characteristic propagation speeds of information in the system~\cite{cevolani2018universal,despres2019twofold}. Spectral properties of collective excitations can also be extracted from the dynamics of correlation functions~\cite{gritsev2007spectroscopy,menu2018quench,schemmer2018monitoring,villa2019unraveling}. 

It has been recently shown that in a generic many-body system the dispersion relation of collective excitations can be obtained by measuring equal-time correlation functions following a global quench~\cite{villa2019unraveling}.
Such global quench spectroscopy relies on the insight that a sudden change of a global external parameter generates an out-of-equilibrium initial state populated by a bunch of low-lying quasiparticle excitations, which spread throughout the system.
Separating out their different contributions, one can reconstruct the excitation spectrum using a space-time Fourier transform of equal-time correlation functions.
Since global quenches preserve translation invariance, the use of non local observables, such as two-point correlation functions, is necessary to obtain the dispersion relation of the elementary excitations in homogeneous many-body systems.
In contrast, local quenches break the translation invariance of the initial state and the dynamics of local observables may be sufficient to reconstruct the elementary excitations of homogeneous systems, in spite of their extended nature.
So far, local spectroscopic techniques have explicitly constructed extended elementary excitations site-by-site~\cite{jurcevic2015spectroscopy}. 

In this paper, we show that the non equilibrium dynamics following a \emph{single} local quench out of the ground state is sufficient to unveil the excitation spectrum of a homogeneous quantum many-body system.
More precisely, we extend quench spectroscopy to the case of sudden \emph{local} quenches, of the type routinely realised in current-generation experiments~\cite{fukuhara2013quantum,fukuhara2013microscopic}.
We show that appropriate choices of the quench and the local observable allow us to unveil the elementary excitation spectrum from the ground state as well as transition energies between excited states. Our approach is benchmarked on a variety of spin and particle models using state-of-the-art numerical approaches. Our main goal is to demonstrate the value of local quench spectroscopy as a new experimental method which can be implemented in a straightforward way. By comparison with established pump-probe techniques, quench spectroscopy offers a simpler experimental protocol: here, the quench plays the role of the pump, immediately generating a superposition of excitations without requiring the fine-tuning of resonant laser beams to address individual excitations directly.

The paper is organised as follows.
After briefly introducing the models we consider~(Sec.~\ref{sec:examples_and_initial_state}), we discuss the local quench spectroscopy approach~(Sec.~\ref{sec:LQS}).
We then discuss how to extract elementary excitation spectra (Sec.~\ref{sec:role_observable_sigma_x}) as well as energy transitions between excited states~ (Sec.~\ref{sec:explanation_branches_QSF_gq}), benchmarking the corresponding approaches on various models.
We finally summarise and discuss the results~(Sec.~\ref{sec:conclusions}).

\section{Models}
\label{sec:examples_and_initial_state}
To illustrate the local quench spectroscopy approach,
we consider three specific lattice models.
We focus on one-dimensional systems at zero temperature for which exact numerical simulations can be efficiently performed, and which are particularly suitable for quantum simulation of strongly-correlated regimes~\cite{giamarchi2004quantum}.
The approach is, however, equally valid in higher dimensions.
The models which we study in this work are routinely realised in experimental platforms such as trapped ions~\cite{jurcevic2015spectroscopy}, Rydberg atoms~\cite{labuhn2016tunable,bernien2017probing,browaeys2020many}, and ultracold atomic gases~\cite{trotzky2008time,cheneau2012light,takasu2020energy}.
All are able to reach the requisite low temperatures and offer a high degree of control, including the single-site or -atom addressability~\cite{sherson2010single,endres2011observation} necessary for the engineering of local quenches.

\subsection{The transverse field Ising chain}
\label{sec:model_TFI}
The first model we consider is the transverse-field Ising (TFI) spin-$1/2$ chain, given by the Hamiltonian
\begin{equation}
\label{eq:hamiltonian_tfi}
\begin{split}
\hat{H}_{\rm{TFI}}&=-J\sum\limits_{R}\hat{\sigma}_{R}^{x}\hat{\sigma}_{R+1}^{x}-h\sum\limits_{R}\hat{\sigma}_{R}^{z},
\end{split}
\end{equation}
where $h$ is the transverse field, $J>0$ the ferromagnetic exchange coupling and $\hat{\sigma}^{\alpha}_{R}$ the Pauli matrix along the axis $\alpha$ at the lattice site $R$. A quantum phase transition at $h=J$ separates a $z$-polarised paramagnetic phase (for $h>J$) from a doubly degenerate $\pm x$-polarised ferromagnetic phase (for $h<J$).
The exact excitation spectrum can be obtained by a mapping to non interacting spinless fermions using the Jordan-Wigner transformation, followed by a Bogoliubov transformation~\cite{franchini2017introduction}, which yields
\begin{equation}
\label{eq:excit_spectrum_TFI}
E_k=2\sqrt{h^{2}+J^{2}-2hJ\cos k},
\end{equation}
where $k$ is the excitation momentum.
In the following discussion, we will restrict ourselves to the $z$-polarised phase in which the elementary excitations are local spin flips~\cite{sachdev1999quantum} and are suitable to local quench spectroscopy. In contrast, the $x$-polarised phase is characterised by topologically non local domain-walls, which are not directly amenable to local quench spectroscopy.
The excitation spectrum in the $x$-polarised phase can, however, be obtained from that of the $z$-polarised phase via self-dual symmetry -- see Appendix \ref{app:non_local_excitations} for details.

\subsection{The Heisenberg chain}
\label{sec:model_Heisenberg}
The second model is the Heisenberg spin-$1/2$ chain, governed by the Hamiltonian
\begin{equation}
\label{eq:hamiltonian_xxx}
\hat{H}_{\rm{H}}=-J\sum_{R}\left(\hat{\sigma}^{x}_{R}\hat{\sigma}^{x}_{R+1}+\hat{\sigma}^{y}_{R}\hat{\sigma}^{y}_{R+1}+\hat{\sigma}^{z}_{R}\hat{\sigma}^{z}_{R+1}\right),
\end{equation}
where $J>0$ is the isotropic ferromagnetic spin exchange coupling between nearest neighbours. The Heisenberg chain is a non trivial interacting model with a highly degenerate ground state in which all spins are aligned in the same but arbitrary direction, irrespective of the value of $J>0$. 
Nevertheless, this model is integrable and can be solved by Bethe ansatz~\cite{bethe1931theorie}.
It yields the excitation spectrum 
\begin{equation}
\label{eq:DR_xxx}
E_{k}=4J(1-\cos k).
\end{equation}
In contrast to the TFI model, however, the real-space representation of the excitations is unknown. We will nonetheless demonstrate that local quench spectroscopy does not require such prior knowledge to probe the spectral properties of the model, as the underlying mechanism is generic. 

\subsection{The Bose-Hubbard chain}
\label{sec:model_BHM}
We finally consider the Bose-Hubbard (BH) chain, given by the Hamiltonian
 \begin{equation}
 \label{eq:BHm}
\hat{H}_{\rm{BH}}=-J \sum_{R} \left(\hat{a}^{\dagger}_{R}\hat{a}_{R+1}+\t{H.c.}\right)+\frac{U}{2}\sum_{R} \hat{n}_{R}(\hat{n}_{R}-1),
\end{equation}
where  $J>0$ is the nearest-neighbour hopping amplitude,
$U>0$ is the on-site interaction energy,
$\hat{a}_{R}^{(\dagger)}$ is the annihilation (creation) operator of a boson at the lattice site $R$,
and $\hat{n}_{R}=\hat{a}^{\dagger}_{R}\hat{a}_{R}$ is the corresponding occupation number. 
At temperature $T=0$, the model contains two phases separated by a quantum phase transition: a Mott insulator (MI) for large $U/J$ and integer filling, and a superfluid (SF) otherwise.
The average filling is denoted $\bar{n}=(1/L)\sum_{R} \langle \hat{n}_R \rangle$.

The excitation spectrum of the model is only known in limiting cases. In the weakly interacting superfluid regime, $\bar{n}\gg U/2J$, the excitations are Bogoliubov quasiparticles~\cite{natu2013dynamics} with the excitation spectrum
  \begin{equation}
  \label{eq:DR_BHm}
E_k=2J\sqrt{2\sin^{2}\left(\tfrac{k}{2}\right)\left[2\sin^{2}\left(\tfrac{k}{2}\right)+\bar{n}\tfrac{U}{J}\right]}.
  \end{equation}
Deep in the Mott phase, the BH chain can be mapped to a fermionic model which admits two types of excitations \cite{barmettler2012propagation}.
Their excitation spectra for $\bar{n}=1$ are
\begin{equation}
\begin{split}
\label{eq:doublon_holon_DR}
E_{k}^{\mp}&=\mp J\cos k +\frac{1}{2}\sqrt{(U-6J\cos k)^{2}+32J^{2}\sin^{2}k},\\
\end{split}
\end{equation}
respectively.
In the strongly interacting limit $U\gg J$, these excitations correspond to doublons and holons.

\section{Local Quench spectroscopy}
\label{sec:LQS}
Let us now discuss the general idea underlying local quench spectroscopy.

\subsection{The quench spectral function}
\label{sec:QSF}
The quench spectral function (QSF) was first introduced in Ref.~\cite{villa2019unraveling} for global quantum quenches. Here we extend it to \textit{local} quenches where translation invariance of the initial state is broken, and the observables are local one-point functions. Following a local quench at time $t=0$, we study the dynamics of a local observable $\hat{O}(\vec{x},t)$ at a given position $\vec{x}$ and time $t>0$, and compute the expectation value
\begin{equation}
\label{eq:observable}
G(\vec{x};t) = \avg{\hat{O}(\vec{x},t)}=\text{Tr}\left[\hat{\rho}_{\textrm{i}}\hat{O}(\vec{x},t)\right],
\end{equation}
where $\hat{\rho}_{\textrm{i}}$ is the density matrix immediately after the quench.
The translation invariance of the initial state is broken, but the Hamiltonian (which is unchanged) remains translationally invariant.
Using the translation operator, we decompose $\hat{O}(\vec{x}) =\e^{-i\hat{\vec{P}} \cdot \vec{x}}\hat{O}(\vec{0})\e^{+i\hat{\vec{P}} \cdot \vec{x}}$, where $\hat{\vec{P}}$ is the total momentum operator and we set $\hbar=1$. In a common eigenbasis of $\hat{H}$ and $\hat{\vec{P}}$, each eigenstate $\vert n\rangle$ has a well defined momentum $\vec{P}_{n}$.
Equation~(\ref{eq:observable}) then reads as
\begin{equation}
\label{eq:SM_QSF_not_simplified}
\begin{split}
G(\vec{x};t)&=\sum\limits_{n,n'}\rho_{\mathrm{i}}^{n'n}\e^{i(E_{n}-E_{n'})t}\e^{i(\vec{P}_{n'}-\vec{P}_{n})\vec{x}}\bra{n}\hat{O}\ket{n'},
\end{split}
\end{equation}
where $\hat{O} \equiv  \hat{O}(\vec{0},0)$, and $\rho_{\mathrm{i}}^{n'n}$ is the initial density matrix coherence between the states $\ket{n'}$ and $\ket{n}$.
The QSF is defined by taking the space-time Fourier transform of Eq.~\eqref{eq:SM_QSF_not_simplified}:
\begin{equation}
\label{eq:QSF_general_formula}
\begin{split}
G(\vec{k},\omega)=&\int\d \vec{x} \, \d t\, \e^{-i(\vec{k}\vec{x}-\omega t)} \, G(\vec{x};t)\\
=&~(2\pi)^{D+1}\sum\limits_{n,n'}\rho_{\mathrm{i}}^{n'n}\bra{n}\hat{O}\ket{n'}\\
&\quad\times\delta(E_{n'}-E_{n} - \omega)\delta(\vec{P}_{n'}-\vec{P}_{n}- \vec{k}),
\end{split}
\end{equation}
where $D$ is the spatial dimension.
Equation~\eqref{eq:QSF_general_formula} will be the basis of the following analysis.

At this point it is worth comparing the QSF for local quenches discussed here with the case of global quenches.
The key element in spectroscopy is the emergence of frequency and momentum resonances corresponding to elementary excitations between eigenstates, $n \leftrightarrow n'$.
For a local quench, the translation invariance of the initial state is explicitly broken by the quench, and local (one-point) observables $\hat{O}(\vec{x},t)$ are sufficient to produce such resonances at $\omega = E_{n'}-E_{n}$ at $\vec{k}=\vec{P}_{n'}-\vec{P}_{n}$, corresponding to the Dirac functions in Eq.~(\ref{eq:QSF_general_formula}).
In contrast, in the case of a global quench, the initial state is translationally invariant and it is necessary to use non local observables, for instance two-point correlation functions~\cite{villa2019unraveling}.
For more details, see Appendix \ref{app:differences_with_global_quench}.

In contrast with standard spectroscopic techniques, such as those probing the dynamical structure factor, local quench spectroscopy does not require the measurement of a two-time correlation function, but instead only requires measurement of a single local observable, which is both numerically and experimentally advantageous. For further details on the fundamental differences between quench spectroscopy and established pump-probe methods, see the extended discussion in Ref.~\cite{villa2019unraveling}.

\subsection{Local quench spectroscopy protocol}
\label{sec.local}
The general strategy of local quench spectroscopy directly follows from Eq.~\eqref{eq:QSF_general_formula}, and from here we will focus on weak quenches which generate only low-energy excited states.
The local quench protocol we will use takes the same general form for each of the models we consider. We first initialise the system in the ground state, and then we apply a local operator $\hat{L}$ to a single lattice site in the center of the chain.
The key point is that because the Hamiltonian is translationally invariant,
the eigenstates have a well-defined momentum and extend over the full system.
The action of the local operator generates an inhomogeneous state that is thus a superposition of all low-lying excited states, such that the initial state may be written $\ket{\psi_{\rm{i}}} = \hat{L} \ket{0} = \sum_m c_m \ket{m}$. To probe the excitation spectrum from the ground state, two criteria must be met.

First, we need that the selection rules for energy and momentum in Eq.~\eqref{eq:QSF_general_formula} pick out the transition between the ground state and
the excited states. 
Specifically, if $\ket{n}=\ket{0}$ is the ground state so that $\vec{P}_{n}=0$ and $E_{n}=0$, this means that for each eigenstate $\ket{n'}$, the QSF $G(\vec{k},\omega)$ displays a peak at $\vec{k}=\vec{P}_{n'}$ and $\omega=E_{n'}$. (Alternatively, one can exchange the roles of $n$ and $n'$.) The choice of the observable is thus crucial in probing the transitions required to reconstruct the spectrum.

Secondly, the density matrix coherences between the ground state and the excited states, $\rho_{\mathrm{i}}^{n'0}$ (or $\rho_{\mathrm{i}}^{0n}$) must also be non zero. This is realised by a quench such that the initial state has a non zero overlap with both the ground state and the excited states.

One possible choice of observable required to probe the spectrum can be inferred from a simple argument.
We first initialise the system in the ground state. We then apply a local operator $\hat{L} $ such that the post-quench state is $\ket{\psi_{\rm{i}}} = \hat{L} \ket{0}$ and we require the following two conditions to hold:
\begin{align}
i)& \quad \braket{0 | \hat{O} | n'} \neq 0, \label{eq.cond1} \\
ii)& \quad \rho_{\mathrm{i}}^{n'0} = \braket{ n' | \psi_{\rm{i}}} \braket{\psi_{\rm{i}} | 0}  = \braket{ n' |\hat{L} |  0} \braket{0 | \hat{L}^{\dagger} | 0} \neq 0. \label{eq.cond2}
\end{align}
Taking for simplicity the case where the local operator is unitary ($\hat{L}^{\dagger}=\hat{L}^{-1}$), we may write the vacuum as $\ket{0} = \hat{L}^{\dagger} \ket{\psi_{\rm{i}}} =  \sum_m c_m \hat{L}^{\dagger}  \ket{m}$. Inserting this into Eq.~\eqref{eq.cond1} gives
\begin{align}
\braket{0 | \hat{O} | n'}  = \sum_m c_m^{\star} \braket{m | \hat{L}\hat{O} | n'} \neq 0. \label{eq.L}
\end{align}
One suitable -- although not unique -- choice of operator is $\hat{O} = \hat{L}^{\dagger}$, i.e.\ the observable $\hat{O}$ corresponds to an operator which `reverses' the quench we perform with the local operator $\hat{L}$.

This mechanism of `reversing the quench' is intuitive and generic, and does not require any prior knowledge of the form of the excitations. The quench itself generates the excitations, and a correctly chosen observable allows their properties to be measured.
In situations where the quench is not described by the action of a unitary operator, one should substitute $\hat{L}\mapsto(\hat{L}^{-1})^{\dagger}$ in Eq.~\eqref{eq.L}, making the choice of $\hat{O}$ less intuitive. We will however demonstrate that the picture of the observable `reversing' the quench still holds with only minor modifications. 

The remaining challenge is to ensure that the density matrix coherences $\rho_{\mathrm{i}}^{n'0}$ in Eq.~(\ref{eq.cond2}) are non vanishing.
This condition states that, as in any spectroscopic technique, the probe must couple the ground state to the targeted excitations. Here the quench itself plays the role of the pump and a coherent superposition must be created.
This can be accomplished by an operator of the form
$\hat{L} \propto \hat{\mathbbm{1}} + \hat{P}$,
where the operator $\hat{P}$ couples the ground state to the excited states $\ket{n'}$ we want to probe, so that $\braket{0 | \hat{L}^{\dagger} | 0} \propto 1$ and $\braket{ n' |\hat{L} |  0} \propto \braket{ n' |\hat{P} |  0} \neq 0$.
As we will see, this form is exactly satisfied in spin systems by the rotation operator.

\subsection{Local quenches}
We now discuss experimentally realistic quenches suitable for local quench spectroscopy, applicable to spin and particle models.

\begin{figure}
\centering
\includegraphics[width=0.9\columnwidth]{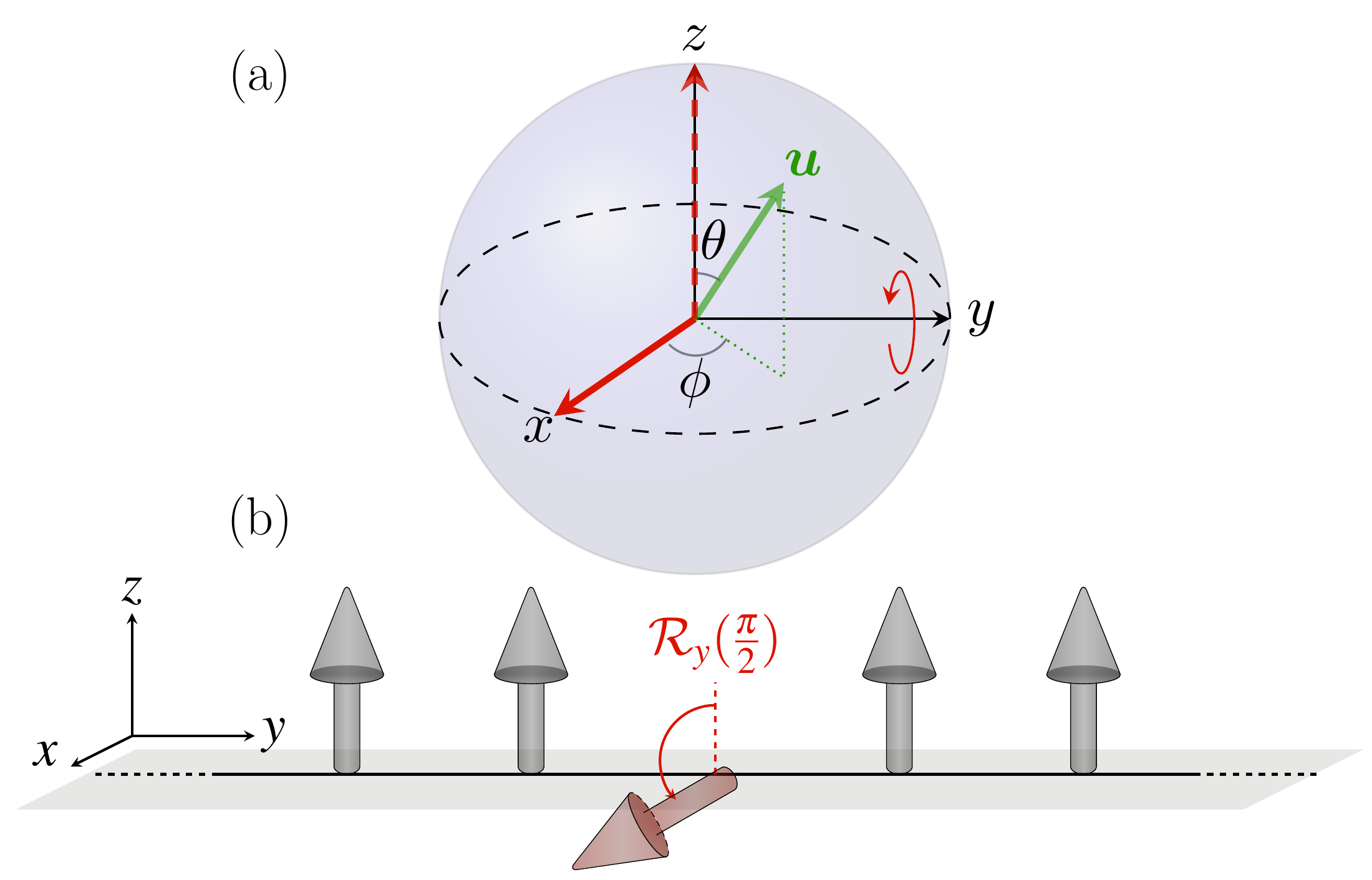}
\caption{(a) Parametrisation $(\theta,\phi)$ of an arbitrary spin-$1/2$ state $\vec{u}$ (green) on the Bloch sphere. (b) Example of a local quench for a spin model. Initially aligned along $z$, the central spin (red arrow) is rotated around the axis $y$ by an angle $\pi/2$ to point along $x$.}
\label{fig:bloch_sphere}
\end{figure}

\paragraph{Spin Models.~--~}
\label{sec:quench_protocol_spins}
For the spin-$1/2$ models,
the local operation applied to the ground state is a rotation
of a spin at some site $\mc{S}$, see Fig.~\ref{fig:bloch_sphere}a.
The spin rotation operator of an angle $\alpha$ around the axis $\vec{n}$ is given by
\begin{align}
\hat{\mathcal{R}}_{{\bf n}}(\alpha) &= \textrm{e}^{-i \frac{\alpha}{2} {\bf n} \cdot \hat{\boldsymbol \sigma}} = \cos \left ( \tfrac{\alpha}{2} \right) \hat{\mathbbm{1}}  - i \sin \left ( \tfrac{\alpha}{2} \right) {\bf n} \cdot \hat{\boldsymbol{\sigma}}.
\end{align}
 As long as the rotation angle $\alpha \neq \pi$, this form satisfies both Eqs.~\eqref{eq.cond1} and \eqref{eq.cond2}.
In both the TFI and Heisenberg models, we rotate the central spin around the axis $y$ by an angle $\theta=\pi/2$. 
In the TFI model, for $h \gg J$, the ground state is strongly polarised along $z$ and thus very close to the product state $\ket{\psi_{\rm{GS}}}\simeq\ket{\uparrow}^{\otimes L}$, where $L$ is the spin chain length.
The after-quench state is given by
  \begin{equation}
  \label{eq:spins_initial_state}
  \begin{split}
  \ket{\psi_{\rm{i}}}\simeq\ket{\uparrow}^{\otimes \mc{S}-1}\otimes\left[\cos(\tfrac{\theta}{2})\ket{\uparrow}_{\mc{S}}+\sin(\tfrac{\theta}{2})\ket{\downarrow}_{\mc{S}}\right] \otimes\ket{\uparrow}^{\otimes L-\mc{S}},
  \end{split}
  \end{equation}  
see Fig.~\ref{fig:bloch_sphere}b. The elementary excitations of this model are spin flips, and so rotating the central spin by $\pi/2$ means that we have prepared the system in a coherent superposition between the $z$-polarised ground state and a state with a single spin flip.

In the Heisenberg model, the ground state is highly degenerate due to global rotational invariance. In our numerical simulations we lift this degeneracy by applying a small magnetic field to the first site of the chain. In the numerics, the field is aligned along the $z$ axis with strength $h_z = - 10^{-2}J$. As with the TFI model, the ground state is thus also $z$-polarised, and so the post-quench initial state is also approximately given by Eq.~\eqref{eq:spins_initial_state}.
 
\paragraph{Bose-Hubbard model.~--~} 
\label{sec:quench_protocol_BH}
In the BH model, local quenches consist of the introduction of defects on a single site. In the superfluid we start from the ground state and then remove all particles in the center. In the Mott insulating phase, we also start from the ground state (close to the homogeneous state with $\bar{n}$ particles per site when $U/J\gg 1$), and remove or add one particle in the center.
In contrast to the spin models, these quenches are non unitary and do not conserve the total particle number.

\subsection{Numerical simulations}
\label{sec:numerics}

One-dimensional quantum systems can be efficiently simulated using numerically exact tensor network methods. Here, we use density matrix renormalisation group (DMRG) to initialise our models in their pre-quench equilibrium ground state.
We then apply a time-dependent variational principle (TDVP) algorithm to compute the time-evolution following the local quench.
DMRG is a well-established variational technique for the study of one-dimensional quantum systems~\cite{white1993density,schollwock2005density}, able to efficiently compute ground state properties of quantum many-body systems.
TDVP is a dynamical variational procedure strongly related to the DMRG algorithm \cite{haegeman2011time,haegeman2016unifying}, which allows efficient simulation of the non equilibrium dynamics of the systems we consider. Both techniques in their modern forms are typically understood in the framework of matrix product states and tensor networks, and we refer the reader to Refs.~\cite{schollwock2011density,vanderstraeten+19} for further details.

In all of the following, we use a maximum bond dimension $\chi=256$ and simulate lattice chains of size $L=47$, comparable with state-of-the-art experiments using ultracold atomic gases and trapped ion platforms. We have ensured convergence of our numerical results by benchmarking with a variety of different bond dimensions and system sizes. In all cases, we use open boundary conditions for the numerical simulations. We choose to make our local quench in the center of the chain to  minimise boundary effects. We have verified that the precise lattice site used does not affect the results.

\section{Extracting excitation spectra}
\label{sec:role_observable_sigma_x}
We now discuss the implementation of local quench spectroscopy to probe elementary excitation spectra.

\subsection{Analytical insight}
\label{sec:spectrum_analytics}

In the following, we will focus on weak quenches which generate low-energy excited states containing at most a single quasiparticle~\cite{wolfle2018quasiparticles}, such that $\ket{n'}=\hat{\gamma}^{\dagger}_{\vec{q}}\ket{0}$ with $\hat{\gamma}_{\vec{q}}^{\dagger}$ the creation operator of a quasiparticle with momentum $\vec{q}$.
In all of the following $\vec{q}=\vec{0}$ stands for the ground state, \ie the vacuum of quasiparticles. The quasiparticles can be either fermionic or bosonic and satisfy the (anti) commutation relations $[\hat{\gamma}_{p},\hat{\gamma}_{p'}^{\dagger}]_{\pm}=\delta_{p,p'}$.
The momentum selection rule in Eq.~\eqref{eq:QSF_general_formula} imposes $\vec{P}_{n}=\vec{q}-\vec{k}$,
and we find
\begin{equation}
\label{eq:QSF_weak_local_quench_hypothesis}
\begin{split}
G(\vec{k};\omega)&=2\pi\int\rho_{\text{i}}^{\vec{q};\vec{q}-\vec{k}}\bra{\vec{q}-\vec{k}}\hat{O}\ket{\vec{q}}\delta(E_{\vec{q}-\vec{k}}-E_{\vec{q}}+\omega)\d \vec{q},\\
\end{split}
\end{equation}
where we moved to the thermodynamic limit and replaced the sum with an integral.
The QSF given by Eq.~\eqref{eq:QSF_weak_local_quench_hypothesis} shows a resonance in $\omega$ for the transitions between the quasiparticle of momenta $\vec{q}-\vec{k}$ and $\vec{q}$. For the sake of simplicity, we restrict the presentation to the case where a single type of quasiparticle is generated by the quench but the discussion can be easily generalised to systems with multiple different types of quasiparticles.

To probe the excitation spectrum, we need an observable $\hat{O}$ that creates or annihilates an excitation, hence coupling the ground state with the lowest excited state manifold. It follows that $\hat{O}$ should not conserve the number of quasiparticles. The simplest form which satisfies this requirement is $\hat{O}=\sum_{\vec{p}}A_{\vec{p}}\hat{\gamma}_{\vec{p}}+\text{H.c.}$, where $A_{\vec{p}}$ is an arbitrary function of the momentum $\vec{p}$ and we write its complex conjugate $\bar{A}_{\vec{p}}$.
The coupling term
$\bra{\vec{q}-\vec{k}}\hat{O}\ket{\vec{q}}$ in Eq.~\eqref{eq:QSF_weak_local_quench_hypothesis} is then non zero in two cases only. Either $\vec{q}=\vec{0}$, and then 
\begin{equation}
\bra{\vec{q}-\vec{k}}\hat{O}\ket{\vec{q}}=\sum_{\vec{p}}\bar{A}_{\vec{p}}\bra{0}\hat{\gamma}_{-\vec{k}}\hat{\gamma}^{\dagger}_{\vec{p}}\ket{0}=\bar{A}_{-\vec{k}}
\end{equation}
or $\vec{q}=\vec{k}$ and then 
\begin{equation}
\bra{\vec{q}-\vec{k}}\hat{O}\ket{\vec{q}}=\sum_{\vec{p}}A_{\vec{p}}\bra{0}\hat{\gamma}_{\vec{p}}\hat{\gamma}^{\dagger}_{\vec{q}}\ket{0}=A_{\vec{k}}.
\end{equation}
The QSF in Eq.~\eqref{eq:QSF_weak_local_quench_hypothesis} thus reduces to
\begin{equation}
\label{eq:QSF_spectrum_analytics}
\begin{split}
G(\vec{k};\omega)&=2\pi\left[A_{\vec{k}}\,\rho_{\text{i}}^{\vec{k};0}\delta(\omega-E_{\vec{k}})+\bar{A}_{-\vec{k}}\,\rho_{\text{i}}^{0;-\vec{k}}\delta(\omega+E_{-\vec{k}})\right].
\end{split}
\end{equation}
For each momentum $\vec{k}$, the Dirac distributions
induce a resonance at $\omega = \pm E_{\pm \vec{k}}$.
The QSF thus yields the dispersion relation of the first excited state manifold, $\vec{k} \rightarrow E_{\pm \vec{k}}$. If the system admits different types of quasiparticles, Eq.~\eqref{eq:QSF_spectrum_analytics} should be modified to include a sum over all types of excitations $r$ and each excitation branch is probed by the QSF.

\subsection{Numerical results}
We now demonstrate the local quench spectroscopy approach, using the models introduced in Secs.~\ref{sec:model_TFI}, \ref{sec:model_Heisenberg}, and \ref{sec:model_BHM}.
The resolution in frequency of the QSF is inversely proportional to the total evolution time. We restrict our simulations to $tJ=10$, which allows for immediate comparison and implementation in state-of-the-art experiments \cite{trotzky_probing_2012,fukuhara2013microscopic,fukuhara2013quantum,jurcevic2014,jurcevic2015spectroscopy}.
Note that our simulations display small finite size effects in the form of reflections at the boundaries in the real-space data. We find that a small number of reflections are not an issue and do not qualitatively alter the QSF.

\paragraph{TFI chain} 
We first consider the transverse field Ising chain (Sec.~\ref{sec:model_TFI}). We initialise the system in the ground state and rotate the central spin along $y$ by $\theta=\pi/2$ such that the obtained initial state is close to the product state of Eq.~\eqref{eq:spins_initial_state}.

In order for the local quench protocol to probe the excitation spectrum, we must choose a suitable observable, as discussed in Sec.~\ref{sec:spectrum_analytics}. In the present case, we require an observable which connects the ground state with the spin-rotated excited state we prepare as the initial state of our quench.
More formally, since the rotated spin in the center of the chain is a linear combination of a spin up and a spin down, the spin raising operator $\hat{\sigma}^{+}=\hat{\sigma}^{x}+i\hat{\sigma}^{y}$ couples it to the ground state. The lowering spin operator $\hat{\sigma}^{-}=\hat{\sigma}^{x}-i\hat{\sigma}^{y}$ has no effect as the resulting state is orthogonal to the ground state. 
  \begin{figure}[t!]
      \centering
\includegraphics[width=\columnwidth]{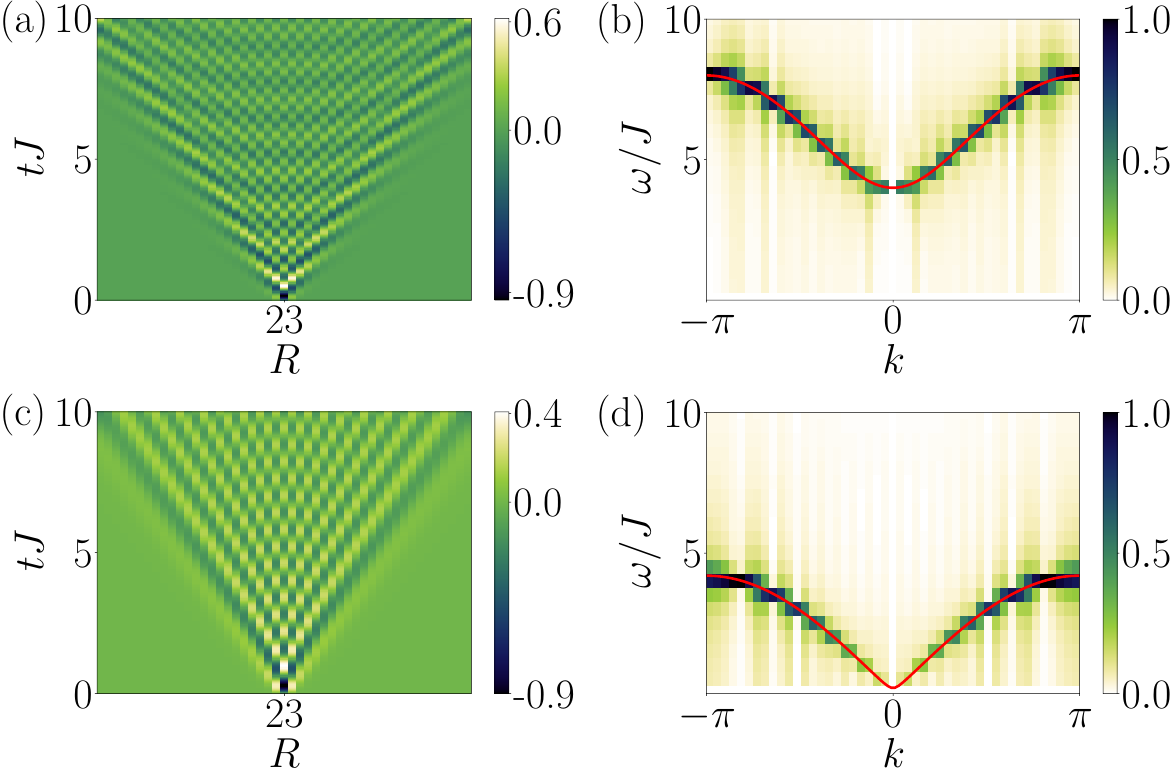}  
    \caption{TDVP simulation of the TFI model with $L=47$ sites. Top: (a)~Dynamics of $\avg{\hat{\sigma}^{y}(R,t)}$ for $h/J=3$, in the $z$-polarised phase. 
    (b)~Corresponding QSF, in excellent agreement with the exact excitation spectrum given by Eq.~\eqref{eq:excit_spectrum_TFI} (red line). Bottom: the same quantities for $h/J=1.1$, close to the critical point.
    Here and in the following, the irrelevant $\omega=0$ component is removed for visibility, and we plot the normalised modulus of the QSF.}
    \label{fig:spectrum_SRTI_along_x}
  \end{figure}
Both observables $\hat{\sigma}^{x}$ or $\hat{\sigma}^{y}$, are linear combinations of the operators $\hat{\sigma}^{\pm}$ and can be used to unveil the spectrum. Equivalently, in the $z$-polarised phase, we can use the Holstein-Primakoff transformation to show that both $\hat{\sigma}^x$ and $\hat{\sigma}^y$ are of the form described in Sec.~\ref{sec:spectrum_analytics}, namely
$\hat{\sigma}^{x}\simeq\hat{\gamma}+\hat{\gamma}^{\dagger}$ and $\hat{\sigma}^{y}\simeq i(\hat{\gamma}-\hat{\gamma}^{\dagger})$, valid for $\avg{\hat{\gamma}^{\dagger}\hat{\gamma}}\ll1$ where $\hat{\gamma}$ and $\hat{\gamma}^{\dagger}$ are the annihilation and creation operators of a spin wave excitation.
In contrast, $\hat{\sigma}^z=2\hat{\gamma}^{\dagger}\hat{\gamma}-1$ is not of this form and does not allow reconstruction of the spectrum, see Sec.~\ref{sec:explanation_branches_QSF_gq}.

Numerical results from TDVP are displayed in Fig.~\ref{fig:spectrum_SRTI_along_x} for $\hat{\sigma}^{y}$ at $h/J=3$ (within the $z$-polarised phase) and $h/J=1.1$ (still in the $z$-polarised phase but close to the critical point). In both cases, the maximum of the QSF modulus coincides with the analytical excitation spectrum~\eqref{eq:excit_spectrum_TFI} (red line).
This excellent agreement with the analytical excitation spectrum remains in the $z$-polarised phase even very close to the critical point $h/J=1$.
For $h/J<1$, corresponding to the $x$-polarised phase, the fundamental excitations are non local domain-walls and it is not possible to find a local operator which couples the ground state to any of the first excited states.
Global quench spectroscopy is therefore preferable in such a case~\cite{villa2019unraveling} -- see Appendix \ref{app:non_local_excitations} for details. In local quench spectroscopy, it is, however, possible to couple the ground state to excited states containing two quasiparticles, in which case instead of a spectrum we see a broad continuum -- see Appendix~\ref{app:additional_signals_QSF_energy_sums}.

\begin{figure}[t!]
      \centering
   \includegraphics[width=\columnwidth]{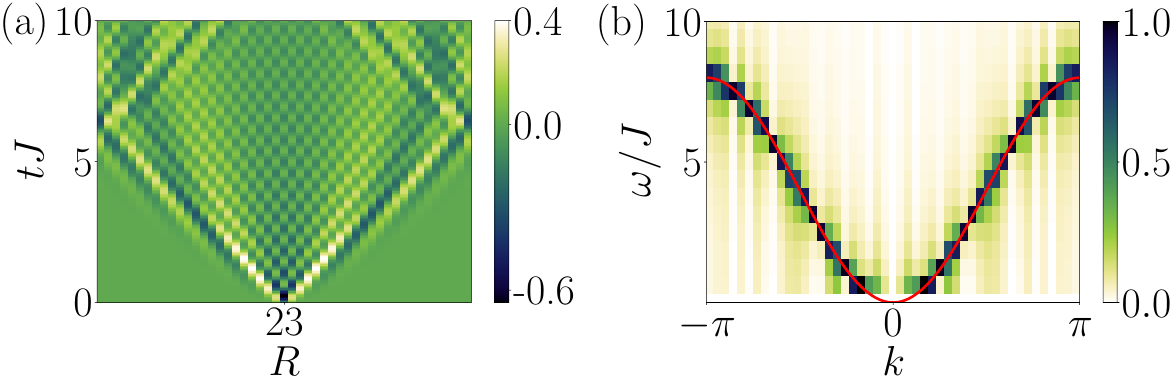}
    \caption{TDVP simulation of the Heisenberg chain with $L=47$ sites. (a)~Dynamics of $\avg{\hat{\sigma}^{y}(R,t)}$ after preparing the system in the ground state and rotating the central spin by $\pi/2$ around the $y$-axis. (b)~Corresponding QSF modulus in excellent agreement with the excitation spectrum predicted by the Bethe ansatz Eq.~\eqref{eq:DR_xxx} (red line). 
}
    \label{fig:heisenberg_XXX}
  \end{figure}

\paragraph{Heisenberg chain} 
We now consider the Heisenberg chain. Contrary to the TFI chain, here the real-space representation of the excitations is unknown, thus there is no {\it a priori} simple choice for the observable. Despite this, using the same quench protocol as for the TFI chain, the numerical results displayed in Fig.~\ref{fig:heisenberg_XXX}, show excellent agreement with the analytical dispersion relation~\eqref{eq:DR_xxx}. This example demonstrates that explicit construction of elementary excitations is unnecessary for the local quench method. Quite generically, a weak local quench populates the low-lying eigenstates of the system. As the quench considered here is the result of a unitary operation, the logic of Sec.~\ref{sec:role_observable_sigma_x} holds exactly, allowing us to obtain the excitation spectrum simply by choosing the observable which `reverses' the quench. The only exception to this rule are cases where the low-lying excited states are protected by a symmetry which cannot be broken by a local operator, e.g. the topologically non local domain-wall excitations present in the TFI model for $h<J$.

\begin{figure}[t]
      \centering
   \includegraphics[width=\columnwidth]{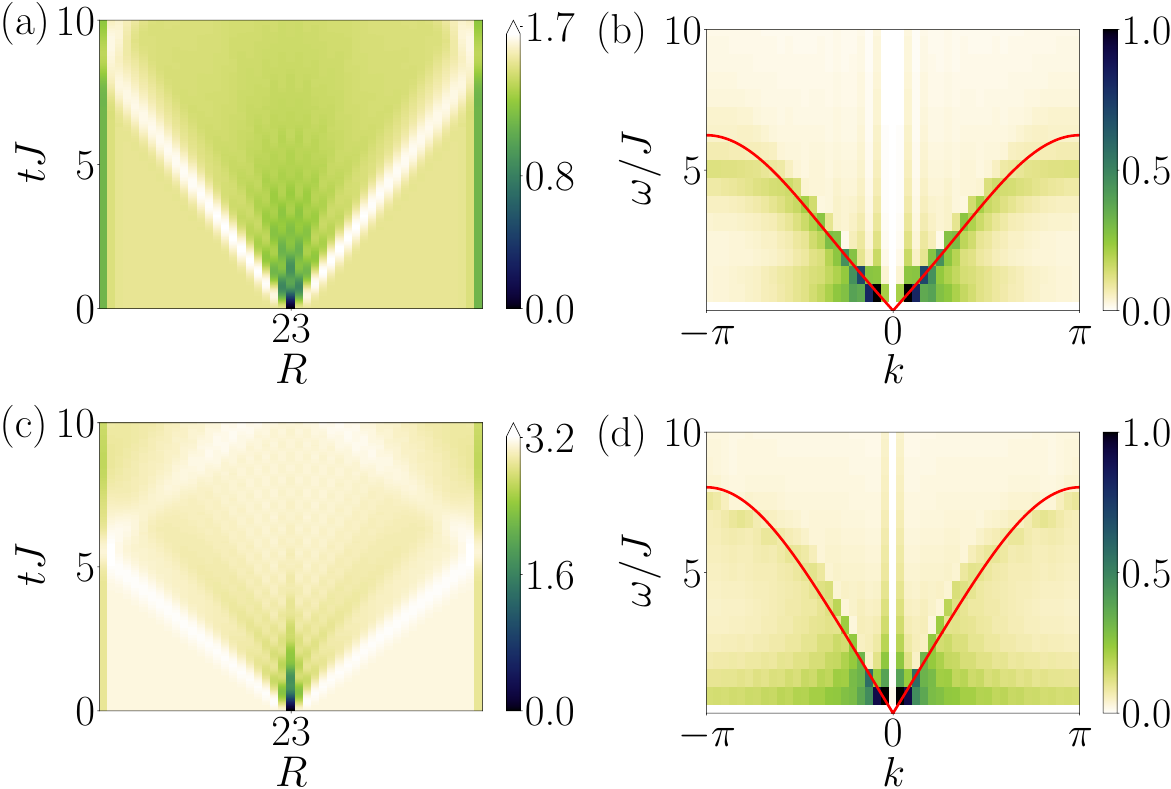}
    \caption{TDVP simulation of the BH chain in the superfluid phase with $U/J=2$ and $L=47$ sites. (a)~Dynamics of $\avg{\hat{n}(R,t)}$ after preparing the system in the ground state with  $\bar{n}\simeq 1.4$, then removing all particles from the center site. (b)~Corresponding QSF modulus in excellent agreement with the excitation spectrum Eq.~\eqref{eq:DR_BHm} (red line). (c)~Dynamics of $\avg{\hat{n}(R,t)}$ after preparing the system in the ground state with  $\bar{n}\simeq 3.0$ and performing the same quench. (d)~The corresponding QSF modulus. The strong signal close to $k=0$ is a signature of quasi long-range order in the 1D superfluid (note that the $k=0$ component itself has been removed for clarity).}
    \label{fig:quench_SF}
  \end{figure}

\paragraph{BH chain}
We finally investigate the Bose-Hubbard chain. We begin by examining a quench in the mean-field superfluid regime, where the excitation spectrum is well approximated by Eq.~\eqref{eq:DR_BHm}. From hydrodynamical formulations, one can expand the density operator as $\hat{n}=n_{0}+\delta\hat{n}$ where $n_0$ is a classical field and $\delta\hat{n}$ is the density fluctuations. In momentum space, $\delta\hat{n}_{k}=A_{k}(\hat{\gamma}_{k}+\hat{\gamma}_{k}^{\dagger})$ (see for instance \cite{pitaevskii_bose-einstein_2016}) such that the density is of the form required to probe the excitation spectrum. After initialising the system in its ground state using DMRG, here we remove all particles on the central lattice site and let the system evolve. Numerical data for two fillings, $\bar{n}\simeq 1.4$ and $\bar{n}\simeq 3.0$, with $U/J=2$ is shown in Fig.~\ref{fig:quench_SF}. The QSF displays clear branches which agree well with Eq.~\eqref{eq:DR_BHm}. 

We now turn to the Mott insulating phase, which contains two different types of excitations propagating with different velocities. 
To understand the nature of the excitations deep in the Mott phase, $U>4(\bar{n}+1)J$, one may restrict the local Hilbert space to $\bar{n}\pm c$ with $c\in\left\{-1,0,1\right\}$, and introduce the fermionic doublons and holons (noted $\hat{c}_{R,\pm}$)  related to the physical particles by $\hat{a}^{\dagger}_{R}=\sqrt{\bar{n}+1}\,Z_{R,+}\hat{c}^{\dagger}_{R,+}+\sqrt{\bar{n}}\,Z_{R,-}\hat{c}_{R,-}$, where $Z_{R,\pm}$ is the Jordan-Wigner phase factor~\cite{barmettler2012propagation}. The excitations are then found by the Bogoliubov transformation
$\hat{\gamma}^{\dagger}_{k,\pm}=u(k)\hat{c}^{\dagger}_{k,\pm}\pm v(k)\hat{c}_{-k,\mp}$. In the strongly interacting regime and to first order in $J/U$, one finds $u(k)=1$ and $v(k)=0$, and the true excitations of the system reduce to pure doublon and holon excitations.

The natural quench we may wish to make in the Mott insulator corresponds to the addition or removal of a single particle. For finite $U/J$, this corresponds to the creation of a linear superposition of both types of excitations. However, this alone is not sufficient for local quench spectroscopy as the resulting state is orthogonal to the ground state and as such it does not satisfy Eq.~\eqref{eq.cond2}. Analogous to the rotation operator used in spin systems, we must prepare a superposition state using a local quench operator of the form $\hat{L} \propto \hat{\mathbbm{1}} + \hat{a}_{L/2}$. The observable we choose to probe is then $\hat{a}^{\dagger}(R,t)$, which again `reverses' the quench, plus its Hermitian conjugate, namely $\hat{O}(R,t) =\hat{a}(R,t)+\hat{a}^{\dagger}(R,t)$.

As this procedure does not conserve the total number of particles in the system, both the preparation of the state and the measurement of the observable will be experimentally challenging. This can be experimentally realised in a double-species Bose-Hubbard model, where we require the bosons to have two different internal hyperfine states (labelled $\ket{\uparrow}$ and $\ket{\downarrow}$) which can be individually addressed. By applying an appropriate laser pulse \cite{duan2003controlling,mandel2003controlled,trotzky2008time,trotzky2010controlling}, it is possible to locally prepare a boson in the required coherent superposition of both hyperfine states. Rather than measuring the bare creation or annihilation operator, the experiment will instead probe the `spin-flip' operator $\hat{a}^{\dagger}_{\downarrow}(R,t) \hat{a}_{\uparrow}(R,t) + \text{H.c.}$ associated to transitions between the two hyperfine states, and which has previously been reconstructed from measurements in Ref. \cite{trotzky_probing_2012}. The two-species Bose-Hubbard model can be mapped onto a pseudo spin Hamiltonian in the strongly-interacting limit \cite{duan2003controlling,altman2003phase,garcia2003spin,isacsson2005superfluid}: this quench again becomes a unitary rotation of the pseudo spin and the required operator again reduces to a transverse magnetisation, precisely as we have previously discussed in the case of spin chains.

\begin{figure}[t]
      \centering
   \includegraphics[width=\columnwidth]{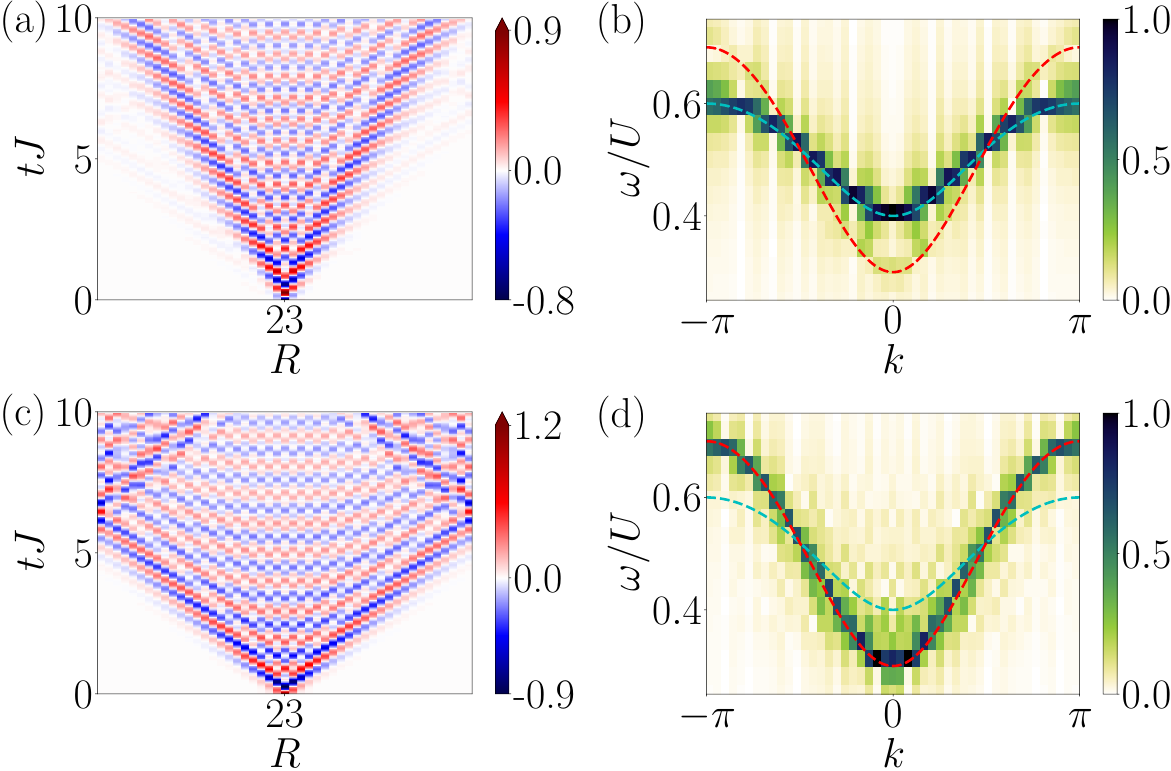}
    \caption{TDVP simulation of the BH chain in the Mott insulator, with $U/J=20$, $\bar{n}=1$. 
    (a) Dynamics of $\avg{(\hat{a}+\hat{a}^{\dagger})(R,t)}$ after the system was prepared with the central site in a coherent superposition of Fock states $(\ket{0}+\ket{1})/\sqrt{2}$. (b) The associated QSF, with the excitation spectra from Eq.~\eqref{eq:doublon_holon_DR} indicated by the cyan and red lines. In this case only one band (close to the holon for $U \gg J$) is visible. (c) Dynamics of  $\avg{(\hat{a}+\hat{a}^{\dagger})(R,t)}$ now with the central site in the coherent superposition of Fock states $(\ket{1}+\ket{2})/\sqrt{2}$. (d) The associated QSF: in this case, only the band close to the doublon is visible (red line).
}
    \label{fig:MI_bplusbdagger}
  \end{figure}

Starting from the ground state with $\bar{n}=1$ and following the procedure above, we prepare the system with the central site in a coherent superposition of Fock states $(\ket{0}+\ket{1})/\sqrt{2}$ and probe the dynamics of $\avg{(\hat{a}+\hat{a}^{\dagger})(R,t)}$. The QSF displays one branch as shown in Fig.~\ref{fig:MI_bplusbdagger}(b) associated to the quasiparticle close to the holon (blue line). If we alternatively {\it add} one particle such that the central site is in the coherent superposition of Fock states $(\ket{1}+\ket{2})/\sqrt{2}$, we can probe the doublon dispersion relation, shown in Fig.~\ref{fig:MI_bplusbdagger}(d). The latter quench may be much more challenging to engineer experimentally, but we include it for completeness.

Finally, we stress that the density operator $\hat{n}(R,t)$ does not couple the ground state to the first excited state manifold, and thus cannot be used to probe the spectrum, though it does allow us to obtain complementary information about the quasiparticles (see Sec.~\ref{sec:explanation_branches_QSF_gq}). Neither can one use $g_1=\avg{\hat{a}^{\dagger}(R,t)\hat{a}(0,t)}$ as was done in Ref.~\cite{villa2019unraveling}, since here for a local quench two-point functions do not allow reconstruction of the excitation spectrum in a simple way (see Appendix \ref{app:differences_with_global_quench}).
 
\section{Local quench spectroscopy for transition energies}
  \label{sec:explanation_branches_QSF_gq}
We now extend the local quench spectroscopy and show that energy differences between excited states within the same excitation manifold can also be probed.
  
  \subsection{Analytical insight}
  \label{sec:analytical_proof_sink/2}
To probe transition energies, we need an observable $\hat{O}$ which conserves the number of quasiparticles, as this restricts the expectation value in the QSF to states within the same manifold. The simplest form of such an observable is $\hat{O}=\sum_{\vec{p},\vec{p'}\neq \vec{0}}B_{\vec{p},\vec{p'}}\hat{\gamma}^{\dagger}_{\vec{p}}\hat{\gamma}_{\vec{p}'}$. The only non zero contributions to the expectation value $\bra{\vec{q}-\vec{k}}\hat{O}\ket{\vec{q}}$ are situations where both $\vec{q}$ and $\vec{q}-\vec{k}$ are non zero, such that
\begin{equation}
\begin{split}
\bra{\vec{q}-\vec{k}}\hat{O}\ket{\vec{q}}&=\bra{\vec{q}-\vec{k}}\sum_{\vec{p},\vec{p}'\neq \vec{0}}B_{\vec{p},\vec{p}'}\hat{\gamma}_{\vec{p}}\hat{\gamma}^{\dagger}_{\vec{p}'}\ket{\vec{q}}\\
&=\sum_{\vec{p},\vec{p}'\neq\vec{0}}B_{\vec{p},\vec{p}'}\delta_{\vec{q},\vec{p}}\delta_{\vec{q}-\vec{k},\vec{p}'}= B_{\vec{q},\vec{q}-\vec{k}},
\end{split}
\end{equation}
where we have discarded an irrelevant $\vec{k}=\vec{0}$ term.
The QSF given by Eq.~\eqref{eq:QSF_weak_local_quench_hypothesis} now reads as
  \begin{equation}
  \label{eq:QSF_energy_differences}
\begin{split}
G(\vec{k};\omega)&= 2\pi \int B_{\vec{q},\vec{q}-\vec{k}}\,\rho_{\text{i}}^{\vec{q};\vec{q}-\vec{k}}\delta(E_{\vec{q}-\vec{k}}-E_{\vec{q}}+\omega)\d\vec{q}\\
\end{split}.
\end{equation}
Here, the delta distribution selects transition energies within the first excited state manifold, rather than coupling to the ground state.
In general, Eq.~\eqref{eq:QSF_energy_differences} does not show any $\delta$-like divergence, owing to the integral over $q$. It can, however, show algebraic divergences along some specific lines $k \rightarrow \omega(k)$. To show this, it is convenient to define the function
\begin{equation}
\label{eq:definition_g}
  \begin{split}
  g_{k}(q,\omega)=E_{q-k}-E_{q}+\omega,
  \end{split}
  \end{equation}  
and call $q^{\star}_{k}(\omega)$ its $\omega$-dependent zeros. As long as $\forall q^{\star}_{k}(\omega)$,$\left.\partial_{q}g_{k}(q)\right|_{q=q^{\star}_{k}(\omega)}\neq 0$, we can rewrite Eq.~\eqref{eq:QSF_energy_differences} as
  \begin{equation}
  \label{eq:formula_delta}
  \begin{split}
G(k;\omega)&=2\pi\sum_{q^{\star}_{k}(\omega)}\left|\partial_{q}g_{k}(q^{\star}_{k}(\omega))\right|^{-1}\\
&\quad\times\int  B_{q,q-k}\,\rho_{\text{i}}^{q;q-k} \delta(q-q^{\star}_{k}(\omega))\d q\\
&=2\pi\sum_{q^{\star}_{k}(\omega)}  B_{q^{\star},q^{\star}-k}\,\rho_{\text{i}}^{q^{\star};q^{\star}-k} \left|\partial_{q}g_{k}(q^{\star}_{k}(\omega))\right|^{-1}.
\end{split}
  \end{equation}
Since the function $g_{k}(q,\omega)$ is in general analytic, a divergence in the QSF can only be observed when $\partial_{q}g_{k}(q^{\star}_{k}(\omega))\rightarrow 0$. 
In some cases, it can lead to a well-defined branch, which is not to be confused with the excitation spectrum, as detailed below.

\paragraph{The special case of a cosine-like dispersion relation}
Let us consider a cosine-like dispersion relation, of the form
\begin{equation} 
\label{eq:generic_form_lattice_DR}
 E_{k}=\Delta - v \cos k,
 \end{equation}
where $\Delta-v$ is the gap, and $v$ the maximum group velocity.
This spectrum indeed changes its convexity exactly in the middle of the half-Brillouin zone, at $k=\pi/2$.
Such a spectrum is relevant to many situations.
For instance, it applies to the TFI model away from criticality ($h \ll J$ or $ h \gg J$), to the Heisenberg model for any $J$,
as well as to doublons and holons in the BH model deep in the MI phase, $U \gg J$.
The energy difference appearing in the selection rule of Eq.~\eqref{eq:QSF_energy_differences} can be rewritten as
  \begin{equation}
  \begin{split}
  E_{q}-E_{q-k}=-2v\sin({k}/{2})\sin({k}/{2}-q).
  \end{split}
  \end{equation}
  The zeros of $g_k(q,\omega)$ are then given by
  \begin{equation}
  g_{k}(q^{\star}_{k}(\omega))=0 \Leftrightarrow \sin({k}/{2})\sin\left[{k}/{2}-q^{\star}_{k}(\omega)\right]=-\frac{\omega}{2v}.
\end{equation}
This equation has solutions provided $\omega\leq |2v\sin({k}/{2})|$,
and we find
 \begin{equation}
 q^{\star}_{k}(\omega)=k/2+\arcsin\left(\frac{\omega/2v}{\sin({k}/{2})}\right),
  \end{equation}
for  $k\neq 0$.
We can then evaluate explicitly
  \begin{equation}
    \begin{split}
\left.\partial_{q}g_{k}(q)\right|_{q=q^{\star}_{k}(\omega)}
&=2v\sin({k}/{2})\sqrt{1-\left(\dfrac{\omega/2v}{\sin({k}/{2})}\right)^{2}}.\\
\end{split}
  \end{equation}
 Using Eq.~\eqref{eq:formula_delta} we finally get
  \begin{equation}
  \label{eq:QSF_divergence_expression_sin}
  G(k;\omega)=\dfrac{ B_{q^{\star}_{k}(\omega),q^{\star}_{k}(\omega)-k}\,\rho_{\text{i}}^{q^{\star}_{k}(\omega);q^{\star}_{k}(\omega)-k} }{\left|2v\sin({k}/{2})\right|\sqrt{1-\left(\dfrac{\omega/2v}{\sin({k}/{2})}\right)^{2}}},
  \end{equation}
Therefore, except for pathological cases where the numerator cancels for all $k$ and $\omega$, the QSF shows a divergence along the line
  \begin{equation}
  \label{eq:dvg_qsf_sink/2}
   \omega = \pm 2v\sin({k}/{2}).
   \end{equation}
This is, as we will see in the following numerical results, exactly the signal that we obtain if observables are chosen so as to conserve the number of quasiparticles.

\begin{figure}[tb!]
\centering
\includegraphics[width=\columnwidth]{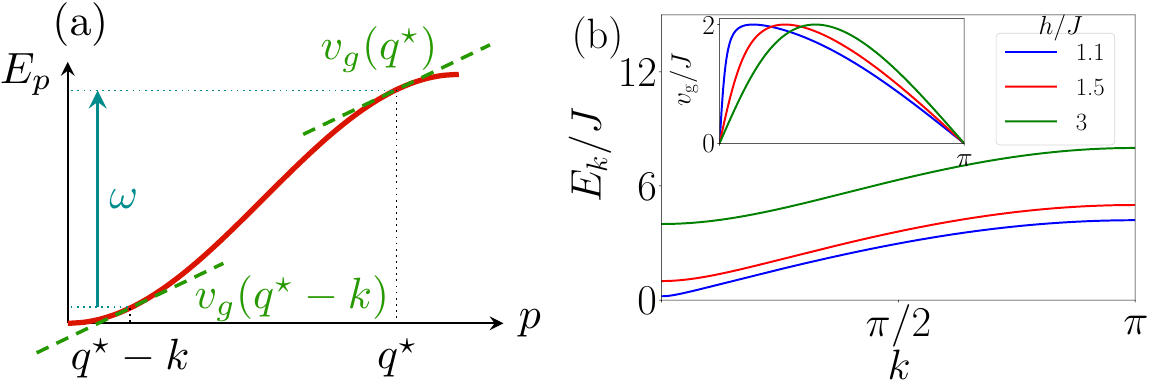}
\caption{(a)~The divergence of the QSF can be interpreted by the interference of two different wavepackets that propagate with the same group velocity. This corresponds to frequencies such that $\omega=E_{q^{\star}}-E_{q^{\star}-k}$, where the notation $q^{\star}$ is short for $q^{\star}_{k}(\omega)$.
(b)~The excitation spectrum of the TFI model strongly deviates from the cosine like spectrum close to the critical point $h/J=1$. Despite this, there always exist two points (except for a single momentum) with different momenta but the same velocity such that Eq.~\eqref{eq:interpretation_DV_QSF_velocities} is satisfied and the divergence in the QSF observed (inset).
}
\label{fig:scheme_velocities}
\end{figure}

\paragraph{Interpretation of the divergence}
From the definition of $g_k$, the zeros of $\partial_{q}g_{k}(q^{\star}_{k}(\omega))$ correspond to the values of $k$ where 
\begin{equation}
\label{eq:interpretation_DV_QSF_velocities}
\begin{split}
v_{\rm{g}}^{\star}(q_{k}^{\star}(\omega)-k)=v_{\rm{g}}^{\star}(q_{k}^{\star}(\omega)),
\end{split}
\end{equation} 
with $v_{\rm{g}}=\partial_{q}E_{q}$ the excitation group velocity.
Hence, the extrema of the QSF correspond to situations where two different wavepackets, formed by quasiparticles created by the quench, propagate with exactly the same group velocity (Fig.~\ref{fig:scheme_velocities}a) and maximise their interference, either constructively or destructively.
A similar phenomenon is known to induce divergences in other contexts, such as in antiferromagnets where caustics in the density of states of two-magnon excitations can be induced by couplings between the ground state and the second excited manifold~\cite{barnes2003Boundaries}. We discuss further features of two-quasiparticle spectra in Appendix.~\ref{app:additional_signals_QSF_energy_sums}.

Equation~\eqref{eq:interpretation_DV_QSF_velocities} can be satisfied for an interval of values of $k$ if and only if $E_{k}$ changes convexity. This condition is far less restrictive than it may appear, as shown on the TFI model for the non trivial situation where $h/J$ is close to 1 [see Fig.~\ref{fig:scheme_velocities}(b)]; there always exists a value $q^{\star}$ for each $k$ such that Eq.~\eqref{eq:interpretation_DV_QSF_velocities} can be satisfied.

   \subsection{Numerical results}
\begin{figure}[t!]
      \centering
    \includegraphics[width=\columnwidth]{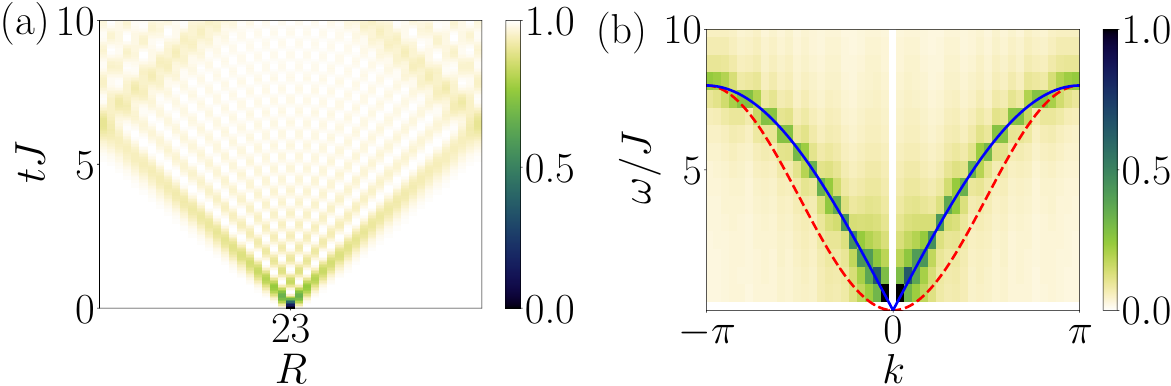}
    \caption{TDVP simulation of the Heisenberg chain. (a)~Dynamics of $\avg{\hat{\sigma}^{z}(R,t)}$.
    (b)~Corresponding QSF modulus, in excellent agreement with the prediction coming from energy differences $8J\sin({k}/{2})$ (blue line).
    The spectrum is displayed as a dashed red line for reference.}
    \label{fig:heisenberg_XXX_Sz}
  \end{figure}

\paragraph{Heisenberg chain}
Here we start with the Heisenberg chain, as the spectrum is exactly of the form of Eq.~\eqref{eq:generic_form_lattice_DR} with $\Delta=v=4J$. 
The local quench procedure is the same as before, and we again initialise the system in the ground state where the central spin is rotated by $\theta=\pi/2$ around $y$,
but here we instead probe $\avg{\hat{\sigma}^{z}(R,t)}$.
Using the Holstein-Primakoff transformation, this operator in terms of quasiparticles reads as $\hat{\sigma}^{z}=2\hat{\gamma}^{\dagger}\hat{\gamma}-\hat{1}$ which conserves the number of quasiparticles and is thus (up to an irrelevant constant term) of the required form discussed in Sec.~\ref{sec:analytical_proof_sink/2}.
The numerical result is shown in Fig.~\ref{fig:heisenberg_XXX_Sz}.
It displays a maximum at $\omega=\pm 8J\sin(k/2)$,
consistent with the prediction of Eq.~\eqref{eq:dvg_qsf_sink/2}.

\paragraph{TFI chain}

\begin{figure}
\centering
\includegraphics[width=\columnwidth]{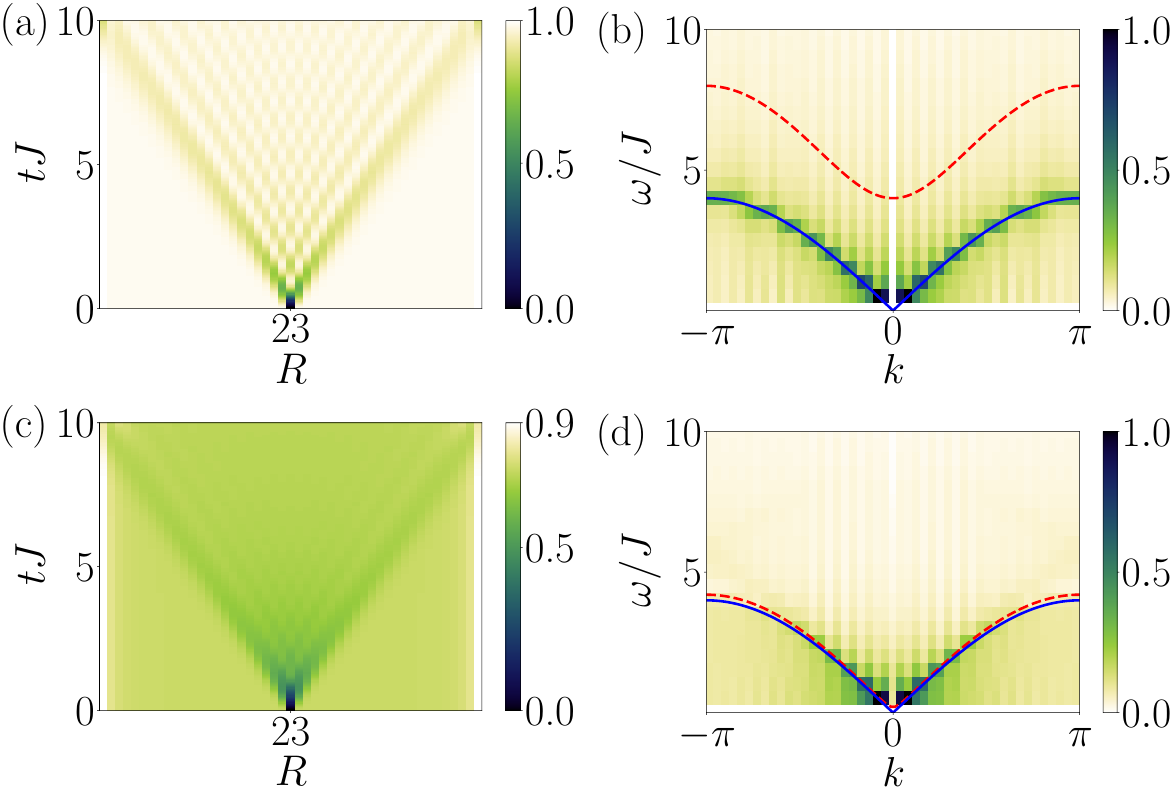}   
\caption{TDVP simulation of the TFI model.
(a)~Dynamics of $\avg{\hat{\sigma}^{z}(R,t)}$ with $h/J=3$.
(b)~Corresponding QSF modulus, in excellent agreement with the predictions coming from energy differences $4J\sin({k}/{2})$ (blue line). As expected, no signal is visible around the excitation spectrum (dashed red). (c)~Dynamics of $\avg{\hat{\sigma}^{z}(R,t)}$ with $h/J=1.1$, close to the critical point. Here, the excitation spectrum is close to, but still distinguishable from, the signal from the energy differences. (d)~Corresponding QSF modulus. Even in this regime, Eq.~\eqref{eq:dvg_qsf_sink/2} (blue line) is a very good fit to the data, with any deviations being below our resolution. }
\label{fig:spectrum_SRTI_along_z}
\end{figure}

We now turn to the TFI chain in the high field regime,  $h \gg J$. Using an expansion to first order in $J/h$,  the excitation spectrum~\eqref{eq:excit_spectrum_TFI} can be cast into the form of Eq.~\eqref{eq:generic_form_lattice_DR} with $\Delta = 2h$ and $v=2J$. We probe $\avg{\hat{\sigma}^{z}(R,t)}$, following the same quench protocol as for the Heisenberg chain. 
The TDVP results are shown in Fig.~\ref{fig:spectrum_SRTI_along_z}a
for the same parameters as in Fig.~\ref{fig:spectrum_SRTI_along_x}a, showing excellent agreement with the prediction of Eq.~(\ref{eq:dvg_qsf_sink/2}), see solid blue line.
In the situation where the excitation spectrum cannot be expanded to give a cosine-like form, for instance as in Eq.~\eqref{eq:excit_spectrum_TFI} close to the critical point, one can still numerically compute the maxima of $E_{q-k}-E_{q}$ with respect to $q$ at fixed $k$: indeed from Eq.~\eqref{eq:interpretation_DV_QSF_velocities} we know that the divergences in the QSF result from such points.
These results are shown in Fig.~\ref{fig:spectrum_SRTI_along_z}(c): remarkably, even close to the critical point, the numerically determined maxima of $E_{q-k}-E_{q}$ are indistinguishable from the prediction of Eq.~\eqref{eq:dvg_qsf_sink/2}, which gives excellent agreement with the numerical data.

\begin{figure}[t]
\centering
\includegraphics[width=\columnwidth]{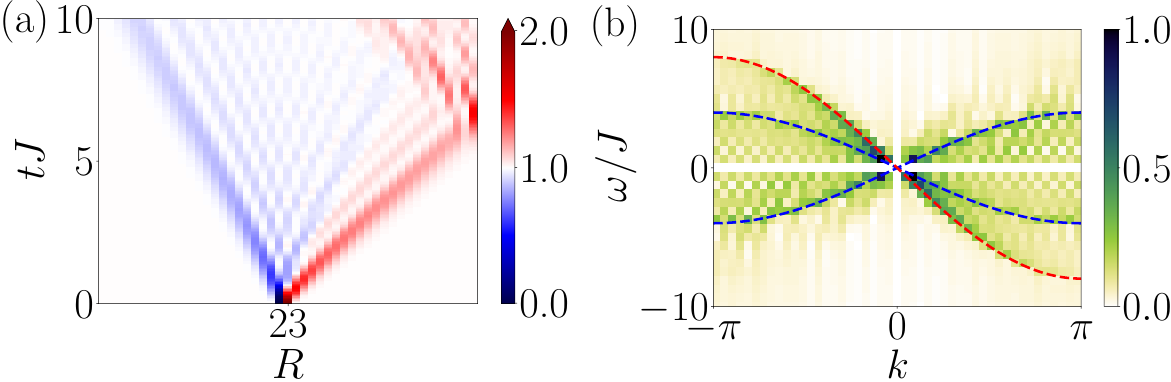}
\caption{TDVP simulation of the BH chain in the Mott insulating phase with $U/J=75$, $\bar{n}=1$. (a)~Dynamics of $\avg{\hat{n}(R,t)}$, where the initial state is the ground state with one particle in the center displaced to the neighbouring site. (b)~Corresponding QSF modulus, in excellent agreement with the predictions coming from energy differences for the doublon $-8J\sin({k}/{2})$ (dashed cyan) and the holon $4J\sin({k}/{2})$ (dashed blue). The $\omega=0$ component is removed for visibility. }
\label{fig:QSF_local_quench_mott_doublon_holon}
\end{figure}

\paragraph{BH chain}
We finally consider the BH model deep in the Mott insulating phase. From the expressions of $\hat{a}_{R}^{(\dagger)}$ given in Sec.~\ref{sec:model_BHM}, the density operator contains terms proportional to $\hat{\gamma}^{\dagger}_{k,\pm}\hat{\gamma}_{k',\pm}$, which are of the required form given in Sec.~\ref{sec:analytical_proof_sink/2}, and can in principle probe both types of excitations. 
Time evolution of the density and the associated QSF for $U/J=75$ and $\bar{n}=1$ are displayed in Fig.~\ref{fig:QSF_local_quench_mott_doublon_holon}. The initial state is the ground state on which a particle-hole pair has been created in the center by moving one particle to the neighbouring site.
The QSF displays several distinct branches which match the energy differences Eq.~\eqref{eq:dvg_qsf_sink/2} for the doublon ($v=4J$) and the holon ($v=2J$), respectively. The doublon propagates twice as fast as the holon due to the Bose enhancement factor~\cite{barmettler2012propagation}, as can be seen by expanding Eq.~\eqref{eq:doublon_holon_DR} to leading order in $J/U$. Note that the two branches for the holon correspond to left- and right-moving signals, respectively, while the single branch for the doublon is for the right-moving signal alone, matching the signal seen in the real-space data in Fig.~\ref{fig:QSF_local_quench_mott_doublon_holon}(a).
  
 \section{Conclusion}
 \label{sec:conclusions}

In this paper, we have extended quench spectroscopy to local quenches.
We have proposed a general scheme able to unveil spectral properties of interacting many-body quantum systems including the elementary excitation spectrum and transition energies between excited states.
Here, we have focused on one-dimensional systems, where quasi-exact tensor network numerical methods are available which allow detailed benchmarking of local quench spectroscopy, and where many experiments are performed~\cite{fukuhara2013microscopic,fukuhara2013quantum,jurcevic2015spectroscopy}.
We expect, however, that the schemes we propose will be equally valid in higher dimensions.
Compared to previous spectroscopic methods which made use of local quenches, our approach requires only a {\it single} local quench. The main advantage is that it does not require any prior knowledge of the physical nature of the excitations, and thus avoids having to explicitly construct the excitation site by site. This offers a considerable simplification over alternative spectroscopic techniques which require fine-tuned addressing of the system in order to create and measure local excitations~\cite{jurcevic2015spectroscopy}.
We have also demonstrated the versatility of the technique, and the range of information that can be accessed using different observables.
On the one hand, we have described a general method to obtain the elementary excitation spectrum by probing an observable which contains terms that reverse the local quench. This choice allows us to induce couplings between the ground state and the elementary excitations, and hence probe the corresponding spectrum, even when the nature of the excitations are not {\it a priori} known.
On the other hand, alternative choices of observable allow us to probe different spectral properties, such as transition energies between excited states. This illustrates the flexibility of local quench spectroscopy, and its capability to extract a great deal of information by probing multiple different aspects of the spectral properties of many-body quantum systems.
From a practical point of view, it is worth pointing out a direct consequence: the observation of a well defined branch does not imply that it coincides with the elementary excitation spectrum. An explicit example of a strongly interacting lattice model was derived analytically together with numerical examples to stress this point.

\acknowledgments
Numerical calculations were performed using HPC ressources from CPHT
and HPC resources from GENCI-CINES (Grant 2019-A0070510300).
We acknowledge the QuSpin package~\cite{weinberg2017quspin,weinberg2019quspin}, which we used in preliminary simulations. The TDVP and DMRG calculations were performed using TenPy~\cite{hauschild2018efficient}.

\appendix  
\section{Difference between global and local quench spectroscopy}
\label{app:differences_with_global_quench}
Here we compare local quench spectroscopy to global quench spectroscopy \cite{villa2019unraveling}. The key difference comes from the breaking of the translational invariance of the initial state, and correspondingly the use of one-point observables instead of two-point correlators.
The derivation of Eq.~\eqref{eq:QSF_general_formula} can be extended to an equal-time correlator between the operators $\hat{O}_{1}(\vec{x},t)$ and $\hat{O}_{2}(\vec{y},t)$ such that
\begin{equation}
\begin{split}
G(\vec{x},\vec{y};t)&=\avg{\hat{O}_{1}(\vec{x},t)\hat{O}_{2}(\vec{y},t)}\\
&=\sum_{n,n',m}\rho_{\rm{i}}^{n'n}\e^{i(E_n -E_{n'})t}\e^{i(\vec{P}_{m}-\vec{P}_n)\vec{x}}\\
&\qquad\times\e^{i(\vec{P}_{n'}-\vec{P}_m)\vec{y}}\bra{n}\hat{O}_{1}\ket{m}\bra{m}\hat{O}_{2}\ket{n'}.
\end{split}
\end{equation}
Assuming translation invariance of the system, it is convenient to change variables by setting $\vec{R}=\vec{x}-\vec{y}$ and $\vec{r}=(\vec{x}+\vec{y})/2$, and
introduce $G(\vec{R};t)=(1/L^D)\int\d\vec{r}\,G(\vec{r},\vec{R};t)$.
Taking the space-time Fourier transform we obtain the QSF
\begin{equation}
\label{eq:G(R,t)}
\begin{split}
&G(\vec{k};\omega)=\int\d\vec{R}\,\d t\,\e^{-i(\vec{k}\vec{R}-\omega t)}G(\vec{R};t)\\
&=\frac{(2\pi)^{D+1}}{L^D}\sum_{n,n',m}\int\left[\d\vec{R} \,\rho_{\rm{i}}^{n'n}(\vec{R})\,\e^{i(\vec{P}_{m}-\vec{P}_{n'}-\vec{k})\vec{R}}\right]\\
&\times\delta(\vec{P}_{n'}-\vec{P}_{n})\delta(E_{n}-E_{n'}+\omega)\bra{n}\hat{O}_{1}\ket{m}\bra{m}\hat{O}_{2}\ket{n'}.
\end{split}
\end{equation}
Below we discuss the consequences of Eq.~\eqref{eq:G(R,t)} for a global and a local quench.
\paragraph{Global quench} 
For a global quench, the initial state is usually taken to be the ground state of the Hamiltonian of the system (for a different set of parameters than the one governing the evolution). Since the system is translationally invariant, so is the ground state.  Therefore $\rho_{\rm{i}}^{n'n}$ becomes independent of $\vec{R}$, which allows us to get another selection rule for the momenta. Equation~\eqref{eq:G(R,t)} now reads as
\begin{equation}
\label{eq:G(R,t)_simplified_global_quench}
\begin{split}
&G(\vec{k};\omega)=\frac{(2\pi)^{2D+1}}{L^D}\sum_{n,n',m}\rho_{\rm{i}}^{n'n}\delta(\vec{P}_{m}-\vec{P}_{n'}-\vec{k})\\
&\times\delta(\vec{P}_{n'}-\vec{P}_{n})\delta(E_{n}-E_{n'}+\omega)\bra{n}\hat{O}_{1}\ket{m}\bra{m}\hat{O}_{2}\ket{n'}
\end{split}
\end{equation}
which is Eq.~(2) used in Ref.~\cite{villa2019unraveling} to extract the excitation spectrum in the case of a weak quench. (Note that the sign of $\omega$ is reversed due to a different Fourier transform convention).
\paragraph{Local quench}
For a local quench, the translation invariance of the initial state is broken, and $\rho_{\rm{i}}^{n'n}$ depends on $R$. The crucial selection rule which links the momentum of $P_{n'}$ with the intermediate state with momentum $P_{m}$ is missing, and the dispersion relation cannot generally be obtained from a two-point function. To circumvent this problem, the QSF has to be defined using one-point observables, as in the main text.

Note that for a global quench, a one-point observable is not able to probe the excitation spectrum because of the translation invariance, which cancels any position dependence. More precisely, the derivation is initially identical to the one presented in the main paper to obtain Eq.~\eqref{eq:SM_QSF_not_simplified}, which we rewrite below for convenience
\begin{equation*}
\begin{split}
G(\vec{x};t)&=\sum\limits_{n,n'}\rho_{\mathrm{i}}^{n'n}\e^{i(E_{n}-E_{n'})t}\e^{i(\vec{P}_{n'}-\vec{P}_{n})\vec{x}}\bra{n}\hat{O}\ket{n'}.
\end{split}
\end{equation*}
Translation invariance of the system imposes that $G(\vec{x};t)=(1/L^{D})\int\d\vec{x}\,G(\vec{x};t)$ is independent of $\vec{x}$. It yields the selection rule $\delta(\vec{P}_{n}-\vec{P}_{n'})$ and the exponential term giving the momentum dependence vanishes.

 \section{Non local excitations in the transverse field Ising chain}
     \label{app:non_local_excitations}
\begin{figure}[b!]
\includegraphics[scale=1]{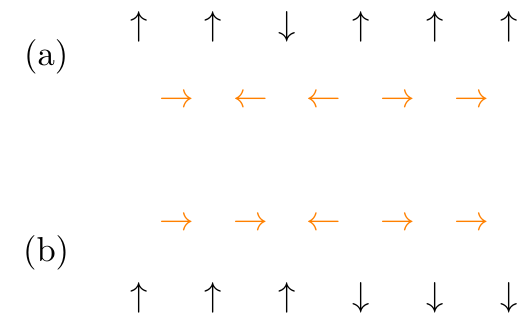}
\caption{(a) Self-duality of the TFI chain, with an original lattice configuration (black) and the reciprocal lattice configuration (orange). With given boundary conditions, the mapping is bijective if the value of the first (or last) spin is fixed in real space (gauge choice).
(b) A spin flip in the reciprocal lattice is equivalent to a domain-wall in the original one.}
\label{fig:mapping-TFI}
\end{figure}

The existence of non local excitations in the TFI chain for $h<J$ can be understood by a self-duality mapping between the two phases.
  The Hamiltonian of the TFI chain is
  \begin{equation}
  \label{eq:hamiltonian_tfi_appendix}
  \begin{split}
  \hat{H}=-J\sum\limits_{j=1}^{N-1}\hat{\sigma}_{j}^{x}\hat{\sigma}_{j+1}^{x}-h\sum\limits_{j=1}^{N}\hat{\sigma}_{j}^{z},
  \end{split}
  \end{equation}
  where the Latin indices $j=1...N$ refer to the lattice spin sites.
We introduce new variables associated to the bonds between spins and denote them by Greek indices $\alpha=1,\cdots,N-1$. On each bond, a spin operator alongside $x$ is introduced with a $+1/2$ value if the bond is ferromagnetic along $z$ ($\uparrow\uparrow$ or $\downarrow\downarrow$) and $-1/2$ for an antiferromagnetic one ($\uparrow\downarrow$ or $\downarrow\uparrow$). An example for a given configuration with $N=6$ is shown in Fig.~\ref{fig:mapping-TFI}(a).   
It corresponds to the introduction of the new spin operators
  \begin{equation}
  \label{eq:mapping_tfi}
  \begin{split}
  \hat{\mu}_{\alpha}^{x}=\prod_{j=1}^{\alpha}\hat{\sigma}_{j}^{z}
  \qquad \textrm{and}\qquad
  \hat{\mu}_{\alpha}^{z}=\hat{\sigma}_{i+1}^{x}\hat{\sigma}_{i}^{x},
  \end{split}
  \end{equation}
which obey the same commutation rules as Pauli matrices. It allows us to rewrite the Hamiltonian \eqref{eq:hamiltonian_tfi_appendix} as
  \begin{equation}
  \begin{split}
  \hat{H}=-h\sum\limits_{\alpha=1}^{N-2}\hat{\mu}_{\alpha}^{x}\hat{\mu}_{\alpha+1}^{x}-J\sum\limits_{\alpha=1}^{N-1}\hat{\mu}_{\alpha}^{z}-h\left(\hat{\sigma}_{1}^{z}+\sigma_{N}^{z}\right)
  \end{split}
  \end{equation}
 Because the last term becomes irrelevant in the thermodynamic limit, the energy spectrum satisfies $E_{k}(\tfrac{h}{J})=\tfrac{h}{J}E_{k}(\tfrac{J}{h})$.
 To overcome the difficulty caused by the doubly degenerate ground state, a gauge may be picked so that the mapping becomes uniquely defined.
 Here we choose the first spin to be always up.

We can now deduce the nature of the excitations for $h<J$ using the mapping.
For $h>J$, excitations are spin flips in the original lattice.
Applying the duality transformation, for $h<J$, the excitations are thus spin flips in the reciprocal lattice. It is equivalent to a domain-wall in the original lattice,~\ie a non local excitation, see Fig.~\ref{fig:mapping-TFI}(b). As there is no local operator which can probe this excitation, and as two-point correlation functions cannot probe the spectrum (as we have shown in Appendix~\ref{app:differences_with_global_quench}) in the case of local quenches, no appropriate local observable can probe the elementary excitations in this phase. We note for completeness that a spin flip in this phase corresponds to two domain-walls, and therefore to the second excited state manifold, which {\it can} be probed using local quench spectroscopy, as we will now discuss.

\begin{figure}[t]
\centering
\includegraphics[width=\columnwidth]{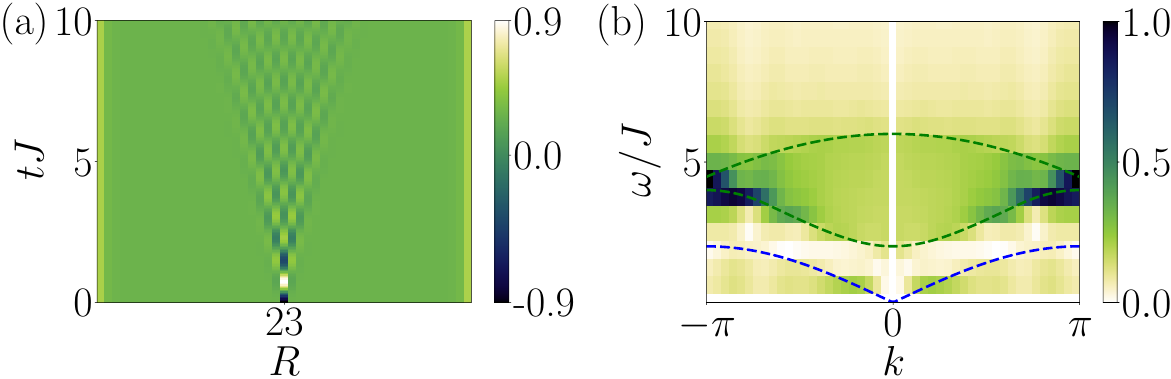}
\caption{TDVP simulations of the TFI model in the $x$-polarised phase for  $h/J=0.5$. (a)~Dynamics of $\avg{\hat{\sigma}^{z}(R,t)}$. The initial state is the ground state and the central spin is rotated by $\theta=\pi/2$ around $y$. (b)~Corresponding QSF modulus. Both the contributions from the energy sums continuum $E_q + E_{q-k}$ (continuum boundaries are represented by the dashed green lines), and the energy differences $E_q - E_{q-k}$ (only the upper boundary is represented by the dotted blue line) can be observed}
\label{fig:spin_chain_bound states_h_0p2}
\end{figure} 

 \section{Additional signals in the QSF}
  \label{app:additional_signals_QSF_energy_sums}
In the main paper we discussed the main features seen in quench spectral functions, in particular the energy spectrum and the sharp branches related to energy differences when the spectrum is cosine-like. For completeness, we comment in this appendix about additional structures which may be observed in the QSF in numerical simulations. We will focus on the ferromagnetic phase, $h<J$, of the TFI chain.
The elementary excitations are non local domain-walls and thus cannot be probed by a local quench (see Appendix \ref{app:non_local_excitations}). However, higher excitations (spin flips) can be unraveled by a local quench as we show below.
\subsection{Analytical insight}  
Starting back from Eq.~\eqref{eq:QSF_general_formula}, we assume that the initial state has a strong overlap with the ground state (say $\ket{n'}$), and account for the contributions in which $\bra{n}$ contains two quasiparticles, with individual momenta $\vec{k_1}$ and $\vec{k_2}$.
We consider for simplicity an observable which can create two quasiparticles of the form $\hat{O}=\sum_{\vec{p},\vec{p'}}C_{\vec{p},\vec{p'}}\hat{\gamma}^{\dagger}_{\vec{p}}\hat{\gamma}^{\dagger}_{\vec{p'}}$ (alternatively one can consider an observable which annihilates two quasiparticles when the role of $n'$ and $n$ is reversed). The momentum selection rule now imposes only $\vec{P}_n=-\vec{k}$, without fixing either $\vec{k_1}$ or $\vec{k_2}$. We choose $\vec{k_1}=-\vec{k}-\vec{q}$ and $\vec{k_2}=\vec{q}$, then
  \begin{equation}
  \begin{split}
  \bra{n}\hat{O}\ket{n'}&=\sum_{\vec{q},\vec{p},\vec{p'}}C_{\vec{p},\vec{p'}}\bra{0}\hat{\gamma}_{-\vec{k}-\vec{q}}\hat{\gamma}_{\vec{q}}\hat{\gamma}^{\dagger}_{\vec{p}}\hat{\gamma}^{\dagger}_{\vec{p'}}\ket{0}\\
  &=\sum_{\vec{q},\vec{p},\vec{p'}} C_{\vec{p},\vec{p'}} \delta_{\vec{p},\vec{q}}\delta_{\vec{p'},-\vec{k}-\vec{q}}\pm(\vec{p}\leftrightarrow \vec{p'})
  \end{split}
  \end{equation}
  (the + is for bosonic quasiparticles, the - for fermionic ones) and $E_{n}=E_{-\vec{k}-\vec{q}}+E_{\vec{q}}$.
  Finally the QSF reads as
  \begin{equation}
  \label{eq:energy_sum}
  \begin{split}
  G(\vec{k};\omega)&=2\pi\int\left( C_{\vec{q},-\vec{k}-\vec{q}}\,\rho_{\rm{i}}^{0;(-\vec{k}-\vec{q},\vec{q})}\pm\rm{sym.}\right)\\
  &\qquad\times \delta(E_{-\vec{k}-\vec{q}}+E_{\vec{q}}+\omega)\d \vec{q}.
  \end{split}
  \end{equation}
[Note that if instead one chooses an observable which annihilates two quasiparticles, with the roles of $n$ and $n'$ reversed, it leads to the symmetric selection rule $(\vec{k},\omega)\rightarrow(-\vec{k},-\omega)$ in energy of the form $\delta(E_{\vec{k}-\vec{q}}+E_{\vec{q}}-\omega)$.]

\subsection{Numerical results}

We start from the DMRG ground state (close to the one where all spins are aligned along the $x$-axis) and rotate the central spin around $y$ by $\pi/2$ such that it points along $-z$. Note that as with the Heisenberg model, here we break the degeneracy of the ferromagnetic phase by applying a small symmetry-breaking field on the first site of the chain with amplitude $h_x = -10^{-2}J$ in the $x$-direction. We compute $\avg{\hat{\sigma}^z(R,t)}$ and display the result in Fig.~\ref{fig:spin_chain_bound states_h_0p2} for $h/J=0.5$.
As expected, in both cases, the QSF computed using TDVP exhibits a continuum.
To compare the numerical results to the prediction, 
we compute the derivative of the function $q\rightarrow E_{-k-q}+E_{q}$ with respect to $q$ to find its extremum, for each value of the parameter $k$, which then allows us to plot the envelope as the dashed green lines in Fig.~\ref{fig:spin_chain_bound states_h_0p2}.
It shows good agreement with the QSF obtained in the numerical simulations.
A continuum originating from the energy differences is also barely visible (dotted blue line). Its origin can be interpreted from the argument detailed in Sec.~\ref{sec:explanation_branches_QSF_gq} here giving a signal close to $2v\sin(k/2)$ with $v=2h$, noticing that the energy spectrum is known for $h<J$ from self-duality (Appendix \ref{app:non_local_excitations}).  

\bibliographystyle{apsrev4-1} 
\bibliography{biblio}

\end{document}